\documentclass[twocolumn,showpacs,preprintnumbers,amsmath,amssymb,10pt,prd]{revtex4}

\usepackage{graphicx}% Include figure files
\usepackage{dcolumn}% Align table columns on decimal point
\usepackage{bm}% bold math
\usepackage{hyperref}
\usepackage{mathrsfs}

\newcommand\be{\begin{equation}}
\newcommand\ee{\end{equation}}
\newcommand\ba{\begin{eqnarray}}
\newcommand\ea{\end{eqnarray}}
\newcommand{\sign}{\mathrm{sign}}
\newcommand{\xmax}{x_{\mathrm{max}}}
\newcommand{\arcsinh}{\mathrm{arcsinh}}

\begin{document}
\title{Duration of inflation and conditions at the bounce as a prediction of effective isotropic loop quantum cosmology}
%\author{Sex, Truck \& Rock'n Roll}

\author{Linda Linsefors}%
 \email{linsefors@lpsc.in2p3.fr}
\affiliation{%
Laboratoire de Physique Subatomique et de Cosmologie, UJF, INPG, CNRS, IN2P3\\
53, avenue des Martyrs, 38026 Grenoble cedex, France
}%

\author{Aurelien Barrau}%
 \email{Aurelien.Barrau@cern.ch}
\affiliation{%
Laboratoire de Physique Subatomique et de Cosmologie, UJF, INPG, CNRS, IN2P3\\
53,avenue des Martyrs, 38026 Grenoble cedex, France
}%

\date{\today}

\begin{abstract}

Loop quantum cosmology with a scalar field is known to be closely linked with an inflationary phase. In this article, we study probabilistic predictions for the duration of slow-roll inflation, by assuming a minimalist massive scalar field as the main content of the universe. The phase of the field in its "prebounce" oscillatory state is taken as a natural random parameter. We find that the probability for a given number of inflationary e-folds is quite sharply peaked around 145, which is consistent with the most
favored minimum values. In this precise sense, a satisfactory inflation is therefore a clear prediction of loop gravity. In addition, we derive an original and stringent upper limit on the Barbero-Immirzi parameter. The general picture of inflation, superinflation, deflation, and superdeflation is also much clarified in the framework of bouncing cosmologies.

\end{abstract}

\pacs{04.60.-m 98.80.Qc}
\keywords{Quantum gravity, quantum cosmology, bouncing cosmology}%Use showkeys class option if keyword

\maketitle

\section{Introduction}

Loop quantum gravity (LQG) is a tentative nonperturbative and background-independent quantization
of general relativity. It uses Ashtekar variables, namely
SU(2) valued connections and conjugate densitized triads. The quantization is 
obtained through holonomies of the connections and fluxes of the densitized triads (see, 
{\it e.g.}, \cite{rovelli1} for an introduction). Basically, loop quantum cosmology (LQC)
is the symmetry reduced version of LQG. In LQC, the big bang is generically replaced by a big bounce due to huge repulsive quantum geometrical effects 
(see, {\it e.g.}, \cite{lqc_review} for a review).

Trying to confront LQG with the real world is a key issue. 
It is not currently possible to compute the cosmological dynamics from the full quantum theory (interesting attempts have recently been presented in \cite{abhay_new}, in particular for perturbations). As in most works on the subject, we will therefore deal with effective equations that are believed to capture the main quantum effects.
Many studies have been devoted to the computation of power spectra and their subsequent comparison with observations (see, {\it e.g.}, \cite{pheno_tensor}). Here, we do not follow this track but, in the spirit of \cite{abhay}, focus instead on only the homogenous part of the Universe, and the probability corresponding to different durations of inflation, within the loop gravity framework. 

Two main LQG corrections are expected when dealing with a
semiclassical approach, as will be the case in this study. The first one comes
from the fact that loop quantization is based on holonomies, {\it i.e.}, exponentials of the
connection rather than direct connection components. The second one arises for inverse
powers of the densitized triad, which, when quantized, becomes an operator with 
zero in its discrete spectrum, thus lacking a direct inverse.  
As the status of "inverse volume" corrections is not fully clear, due to 
the fiducial volume cell dependence, this work focuses on the holonomy term which has a major influence on the background equations.

This paper is organized as follows: 
In section \ref{fram} we present the equations that are the starting point of this paper. 
In section \ref{faces} we explain the different phases of evolution that these equations give rise to. The calculations used in this section can be found in appendix \ref{app}.
In section \ref{pred} we calculate the probability distribution of different solutions, and, in particular the probability distribution of the number of e-folds of slow-roll inflation.
In section \ref{analys} we derive an analytical expression for the most probable value of the number of e-folds of slow-roll inflation.
In section \ref{con} we use the above results to constrain the critical density and the Barbero-Immirzi parameter.

This study is complementary
to the ones performed in \cite{abhay} where the probability distribution was assumed to be flat and defined at the bounce (the first attempts in this direction were performed in \cite{first}). Here, we make a very
different assumption: the phase of the field oscillating in the remote past is considered to be the most natural random variable. As shown in \cite{Corichi:2010zp} the choice of what is a natural measure, and therefore the outcome of these kinds of calculations, can depend heavily on when one decides to define the initial conditions.
Here, we take seriously the meaning of an "initial" condition in a Universe that extends in the past beyond the bounce.
We do not use any heavy machinery
and rely only on very minimalistic hypotheses. Nor do we assume different conditions at the bounce, as in \cite{abhay}, but instead derive them explicitly as predictions of the model.

In the end, we show that, if the critical density is assumed to be a free parameter, a stringent upper limit on the Barbero-Immirzi parameter, $\gamma$, that is the 
free parameter of loop gravity,
can be obtained. This is especially important if, as suggested in \cite{bianchi}, the entropy of black holes can be recovered for any $\gamma$, therefore
leaving its value mostly unconstrained.

The emphasis of this study is put on LQC as this model provides a well-defined framework, with known and controlled equations of motion.  Most results are, however, probably quite generic to bouncing models.

\section{Framework}
\label{fram}

The holonomy-corrected LQC-modified Friedman equation reads as
\begin{equation}
H^2=\frac{\kappa}{3}\rho\left(1-\frac{\rho}{\rho_c}\right) .  \label{Friedman}
\end{equation}
Here we assumed that the Universe is kinetic energy dominated around the bounce so that higher-order terms can be neglected \cite{bojo_prl}. We will see later from the prediction of section \ref{pred} that this is self-consistent.

The main content of the Universe is assumed to be a massive scalar field $\phi$ with mass $m$ fulfilling:
\begin{equation}
\ddot{\phi}+3H\dot{\phi}+m^2\phi=0. \label{eom}
\end{equation} 
This is both the most common and the best motivated (as a scalar can account effectively for many kind of other fundamental contents) choice, allowing for easy comparisons with works carried out in standard cosmology.

We use the critical density, {\it i.e.} density at the bounce, given by \cite{lqc_review}
$
\rho_{\text{c}} = \sqrt{3}m^4_{\text{Pl}}/(32\pi^2\gamma^3) \simeq 0.41 m^4_{\text{Pl}},  
$ 
where $\kappa=8\pi G$ and $\gamma=0.2375$ (except in sections \ref{pred} and \ref{analys} where $\rho_c$ is considered as a free parameter). We use the mass of the scalar field 
$m=1.21\times 10^{-6}$ 
as is favored by observations \cite{linde} (except in section \ref{analys} where $m$ can be taken as a free parameter).

We define the fractions of potential and kinetic energy, normalized to the maximum energy density,
\begin{equation}
x := \frac{m\phi}{\sqrt{2\rho_{\text{c}}}} \quad  \text{and} \quad  y :=\frac{\dot{\phi}}{\sqrt{2\rho_{\text{c}}}}, \label{xy}
\end{equation}
so that
\begin{equation}
\rho=\rho_c\left(x^2+y^2\right).   \label{rhoxy}
\end{equation}
The equations of motion for $x$, $y$, and $\rho$ are
\begin{equation}
\dot{x}=my ~,~  \dot{y}=-mx-3Hy \label{x},
\end{equation}
\begin{equation}
\dot{\rho}=-6H\rho_c y^2 . \label{rhodot}
\end{equation}

\begin{figure*}
	\begin{center}
		\includegraphics[scale=1.1
		]{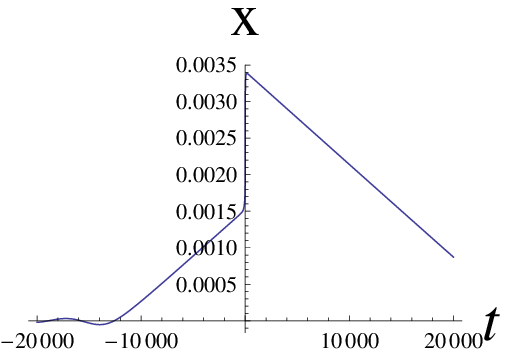} \hfill 
		\includegraphics[scale=1.1]{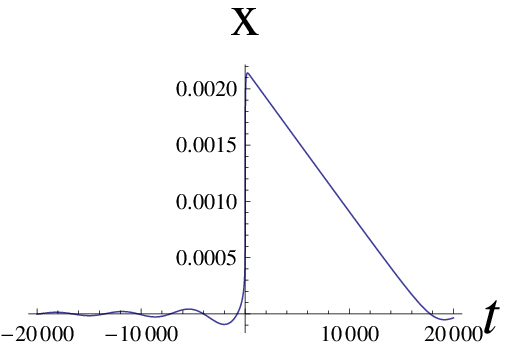}\hfill
		\includegraphics[scale=1.1]{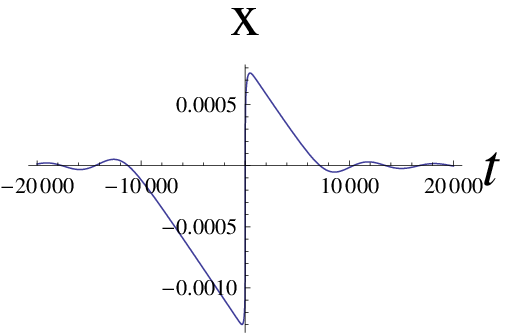}\\
		\includegraphics[scale=1.1]{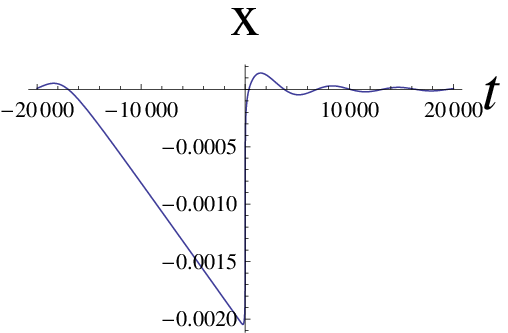}\qquad
		\includegraphics[scale=1.1]{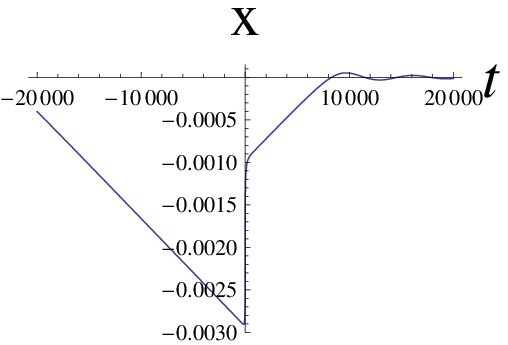}
		\caption{Examples of evolutions of $x$ as a function of time for different solutions.
				The linear increase (decrease) of $|x|$ is the slow-roll deflation (inflation) phase, and the almost vertical increase or decrease of $x$ is the superdeflation, bounce, and superinflation phase. A solution with no deflation at all like in the upper middle plot is by far the most probable.
The mass of the scalar field used here is $m=10^{-3}$ but the features remain true for any mass.}
		\label{fig5}
	\end{center}
\end{figure*}

\section{Phases of the LQC bouncing universe}
\label{faces}

Using Eq. (\ref{Friedman}) and Eqs. (\ref{rhoxy})-(\ref{rhodot}), the evolution of the Universe can be generically described by five phases:\\ 

A. Prebounce oscillations

B. Slow-roll deflation

C. Superdeflation, bounce, and superinflation

D. Slow-roll inflation

E. Post-bounce oscillations\\

Examples of plots of $x$ for different solutions are given in Fig. \ref{fig5}.
They are good indicators of what is happening since one can see here very clearly the differences between the phases of evolution.

We assume that $\rho_c$ is large enough so that $\rho\ll\rho_c$ is always the last of the relevant conditions to be violated before the bounce and the first one to be restored after the bounce. This is in agreement with result from numerical simulations. In the following equations, $t$ is always the cosmic time, but it will be shifted between solutions for the different phases ; for convenience reasons, the convention for the origin of time is not always the same. The exact origin of time is irrelevant for the underlying physics.

The calculations behind the results in this section are presented in appendix \ref{app}.

\subsection{Prebounce oscillations}

This phase is characterized by the fact that $x$ and $y$ are oscillating with vanishing mean values and growing amplitudes.
In this study, we naturally assume this phase to be the initial state of the bouncing Universe. This is of course a hypothesis that can be questioned. The conditions for prebounce oscillations are
\begin{equation}
\rho\ll \rho_c ~,~ H<0 ~,~ H^2\ll m^2. \label{rhoSmalerThanRhoC1}
\end{equation}
The evolution in this phase can be approximated by
\begin{equation}
\rho=\rho_0\left(1-\frac{1}{2}\sqrt{3\kappa\rho_0}\left( t+\frac{1}{2m}\sin(2mt+2\delta)\right)\right)^{-2},\label{prerho}
\end{equation}
\begin{equation}
x=\sqrt{\frac{\rho}{\rho_c}}\sin(mt+\delta)~,~ y=\sqrt{\frac{\rho}{\rho_c}}\cos(mt+\delta).  \label{prex}
\end{equation}
This is stable until $\rho$ grows large enough to violate the last condition.

\subsection{Slow-roll deflation}

Slow-roll deflation is characterized by an almost constant $y$ and a linearly growing $|x|$. The probability of slow-roll deflation is small, since it occurs only if the relation between $x$ and $y$ is very specific at the end of the phase of prebounce oscillations. The conditions for slow-roll deflation are
\begin{equation}
\rho\ll \rho_c ~,~ H<0 ~,~ H^2\gg m^2 ~,~ x^2 \gg y^2. \label{rhoSmalerThanRhoC2}
\end{equation}
In this phase, the equation of motion for $y$ can be approximated by
\begin{equation}
\dot{y}=\sqrt{3\kappa\rho_c}|x|\left(y-\sign(x)\frac{m}{\sqrt{3\kappa\rho_c}}\right).
\label{slowdef}
\end{equation}
The value $y=\sign(x)\frac{m}{\sqrt{3\kappa\rho_c}}$ is an unstable stationary point. The variable $y$ will evolve away from $\sign(x)\frac{m}{\sqrt{3\kappa\rho_c}}$. However, if $y$ starts out very close to $\sign(x)\frac{m}{\sqrt{3\kappa\rho_c}}$, then $\dot{y}\approx 0$ for a while, and this leads to slow-roll deflation.
Slow-roll deflation is in this sense unstable.

\subsection{Superdeflation, bounce and superinflation}

This phase is characterized by a large $|y|$ and a rapidly growing or decreasing $x$ ($y$, and therefore $\dot{x}$, do not change sign during this phase). Superdeflation starts directly after post-bounce oscillations or after slow-roll deflation. The conditions for this phase are
\begin{equation}
H^2\gg m^2~,~y^2\gg x^2.
\label{supercond1}
\end{equation}

The evolution can be approximated by
\begin{equation}
\rho=\rho_c\left(1+3\kappa\rho_c t^2\right)^{-1}~,~y=\pm\left(1+3\kappa\rho_c t^2\right)^{-1/2},
\label{superrho}
\end{equation}
\begin{equation}
x=x_B\pm\frac{m}{\sqrt{3\kappa\rho_c}}\ \arcsinh\left(\sqrt{3\kappa\rho_c}\ t\right),
\label{superx}
\end{equation}
where $t=0$ at the bounce for Eqs. (\ref{superrho})-(\ref{superx}). This phase is stable for $H<0$ but unstable for $H>0$ since, in the later case, $|y|$ is decreasing rapidly and will eventually violate the second condition of Eqs. (\ref{supercond1}).

\subsection{Slow-roll inflation}

Slow-roll inflation happens if the second condition of Eqs. (\ref{supercond1}) is broken before the first one. The conditions for slow-roll inflation are:
\begin{equation}
\rho\ll \rho_c~,~H>0~,~H^2 \gg m^2~,~x^2\gg y^2.  \label{rhoSmalerThanRhoC3}
\end{equation}

In this phase, the equation of motion for $y$ can be approximated by
\begin{equation}
\dot{y}=-\sqrt{3\kappa\rho_c}|x|\left(y+\sign(x)\frac{m}{\sqrt{3\kappa\rho_c}}\right),
\label{slowinf}
\end{equation}
which should be compared with Eq. (\ref{slowdef}). In this case $y=-\sign(x)\frac{m}{\sqrt{3\kappa\rho_c}}$ is an attractor, therefore, slow-roll inflation is stable until one of the two last conditions is violated, which occurs at approximately the same value of $x$ for both conditions.
\footnote{The slow roll parameters are $\epsilon=\eta=\frac{m^2}{\kappa\rho_c x^2}$}

\subsection{Post-bounce oscillations}

The conditions for post-bounce oscillations are
\begin{equation}
\rho\ll \rho_c~,~H>0~,~H^2\ll m^2. \label{rhoSmalerThanRhoC4}
\end{equation}
The evolutions in this phase --corresponding to reheating-- can be approximated by
\begin{equation}
\rho=\rho_0\left(1+\frac{1}{2}\sqrt{3\kappa\rho_0}\left(t+\frac{1}{2m}\sin(2mt+2\delta)\right)\right)^{-2},
\label{postrho}
\end{equation}
together with Eqs. (\ref{prex}).

\section{Numerical predictions}
\label{pred}

\begin{figure}
	\begin{center}
		\includegraphics[width=\columnwidth]{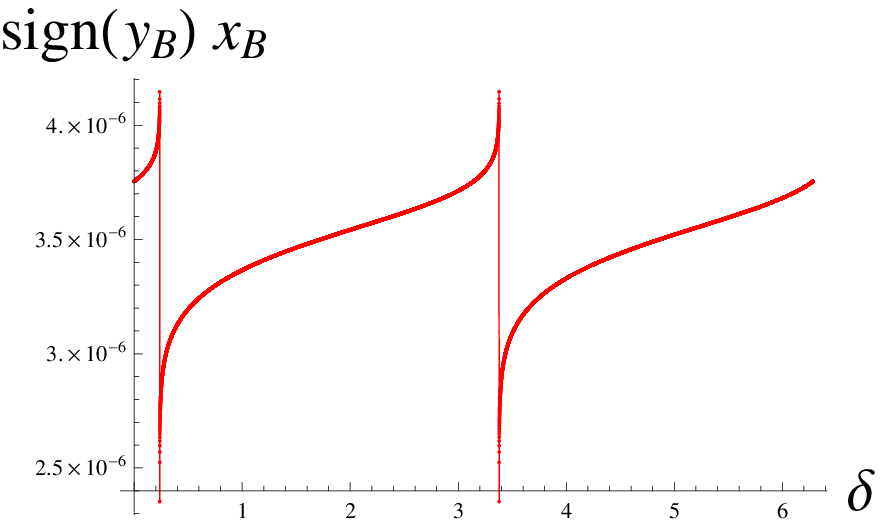}
		\includegraphics[width=\columnwidth]{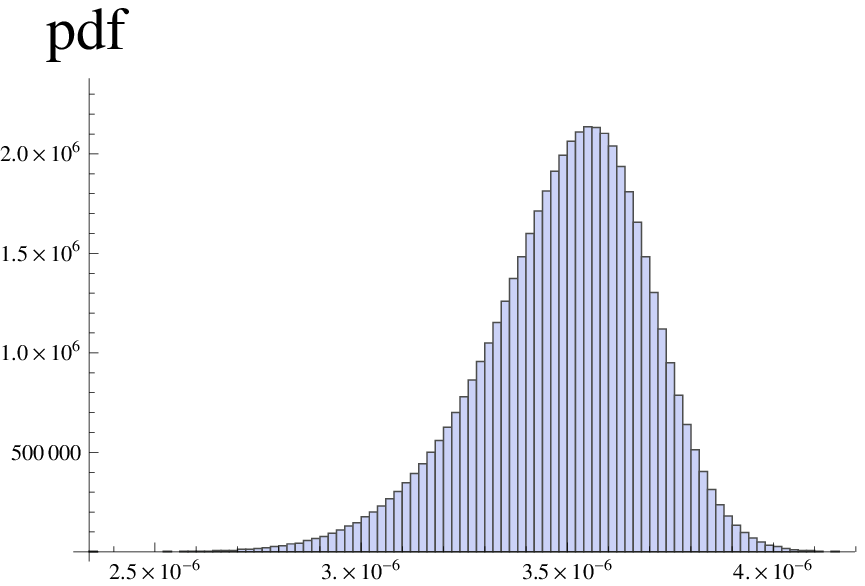}
		\caption{$\sign(y_B)x_B$ as a function of $\delta$ (upper plot) and its probability distribution (lower plot).}
		\label{xb}
	\end{center}
\end{figure}

\begin{figure}
	\begin{center}
		\includegraphics[width=\columnwidth]{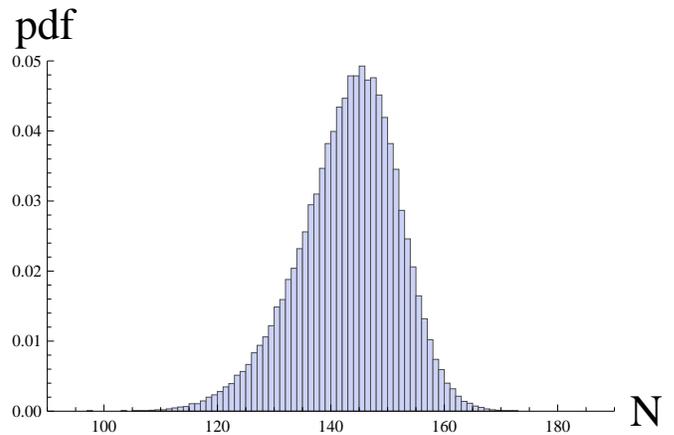}
		\caption{Probability density of the number of e-folds of slow-roll inflation.}
		\label{N}
	\end{center}
\end{figure}
In this section we calculate the probability density function for $x_B$, the square root of the fraction of potential energy at the bounce, and $N$, the number of e-folds of slow-roll inflation. This is done by first finding the most natural initial probability distribution, and then evolving it numerically.

We believe that it is most natural and consistent with the Big Bounce model to set the initial probability distribution in the prebounce oscillation phase.
The evolution of the Universe in this phase is described by Eqs. (\ref{prerho})-(\ref{prex}), with parameters $\rho_0$ and $\delta$. However, the transformation
\begin{equation}
\label{transform1}
\rho_0\rightarrow \rho_1 \\
\end{equation}
corresponds to
\begin{equation}
\label{transform2}
\begin{array}{l}
\delta\rightarrow \delta-\frac{2m}{\sqrt{3\kappa\rho_1}}\left(1-\sqrt{\frac{\rho_1}{\rho_0}}\right),\\
t\rightarrow t+\frac{2}{\sqrt{3\kappa\rho_1}}\left(1-\sqrt{\frac{\rho_1}{\rho_0}}\right),
\end{array}
\end{equation}
and does not therefore generate new solutions. This allows us to take $\delta$ as the only parameter. 

In addition of being the obviously expected distribution for any oscillatory process of this kind, a flat probability for $\delta$ will be preserved over time within the prebounce oscillation phase (under the assumptions given by Eq. (\ref{rhoSmalerThanRhoC1})), making it a very natural choice for initial conditions. By ``preserved over time'' it is meant here that it is preserved by transformations described by Eqs. (\ref{transform1})-(\ref{transform2}). This is not a trivial point as any other probability distribution would be distorted over time, meaning that the final result in the full numerical analysis would depend on the choice of $\rho_0$. It can be noticed that time itself is not a relevant parameter in the numerical analysis: the time it takes for the Universe to evolve from its initial state to the bounce is determined by $\rho_0$.

Starting with a flat probability distribution for $\delta$, and choosing $\rho_0$ so that the solution is initially well approximated by Eqs. (\ref{prerho})-(\ref{prex}), the probability for different values of $x_B$ can be calculated numerically using the full set of Eqs. (\ref{Friedman}) and (\ref{rhoxy})-(\ref{x}). At the bounce the solutions can be parametrized by $x_B$ and $\sign(y)$ ; however only the relative sign is physical. We therefore project the result down to the physically relevant parameters by considering $\sign(y_B)x_B$. The value of $\sign(y_B)x_B$ as a function of $\delta$ and the resulting probability distribution are shown in Fig. \ref{xb}.

In previous works, $\sign(y_B)x_B$ was taken as unknown \cite{abhay}. However, here, we show that it is sharply peaked around $3.55\times 10^{-6}$ (this values scales with $m$ as $m\log\left(\frac{1}{m}\right)$, where we assumed that $m\ll1$ in Plank units). The most likely solutions are exactly those that have no slow-roll deflation. In the tails of the probability spectrum, there are solutions with some slow-roll deflation, but the probability density decreases very rapidly with the length of slow-roll deflation. This result is also expected from the arguments given in section \ref{faces}.

Our result is not symmetric under a time reversal transformation. This is not surprising as we broke the time symmetry of the model by choosing initial conditions outside of the bounce. There is here a clear causal evolution from the past to the future. However, given the same prior (initial) distribution for the post-bounce
oscillation phase and evolving backward, one would of course find that these results hold for the probability of prebounce deflation. 

This result also shows that the bounce is strongly kinetic energy dominated, leading to backreaction effects that can be safely neglected \cite{bojo_prl}. The model is therefore self-consistent.\\

Slow-roll inflation starts when $|x|=\xmax$ where $\xmax\doteq\max_{t>t_B}(|x|)$, which is related to the length of slow-roll inflation by
$
N=\frac{\kappa\rho_c}{2}\left(\frac{\xmax}{m}\right)^2\simeq 5.1\left(\frac{\xmax}{m}\right)^2,
$
where $N$ is the number of e-folds during slow-roll inflation.  The probability density for $N$ is given in Fig. \ref{N}, showing that the model leads to a slow-roll inflation of about 145 e-folds. This becomes an important and clear prediction of effective LQC ; inflation and its duration are not arbitrary in the model.

This prediction is in agreement with observations that require a slow-roll inflation longer than 65 e-folds. Strictly no fine-tuning was required to obtain this result.\\

The old and well-known "measure problem" in cosmology is basically related to the way ignorance should be described. Ignorance means a flat probability distribution function over some natural measure, and the question is the following: what should this measure be ? In the minisuperspace (homogeneous, isotropic and flat) approximation used in this study, the relevant problem is not related with the existence of infinitely many degrees of freedom or with divergent integrals but with the way to chose the significant measure with respect to which the probability distribution is flat. If we assume no knowledge of quantum gravity, it might be a reasonable assumption to chose the "time of ignorance" at  the Planck density, and search for a natural measure of the parameters at that time. However, in this study, we assume that, through the bounce, the Universe is well described by Eqs. (1) and (2). In this approach, our ignorance starts when the mater content begins to be well approximated by the (effective) scalar field (assuming, {\it e.g.}, that the prebounce oscillation phase is created by some inverse reheating process). As we know neither the details of this process --which might very well be purely random-- nor the density at which this occurs, we translate this ignorance as a flat probability distribution for the most natural parameter of this phase. In addition, even if we somehow gain knowledge of the physics governing the "inverse reheating", and even if this this theory predicts a nonflat probability distribution for $\delta$, unless this probability distribution is extremely peaked around the specific value that gives significant slow-roll deflation, our result will hold.

\section{Analytical predictions}
\label{analys}

\begin{figure}
	\begin{center}
		\includegraphics[width=\columnwidth]{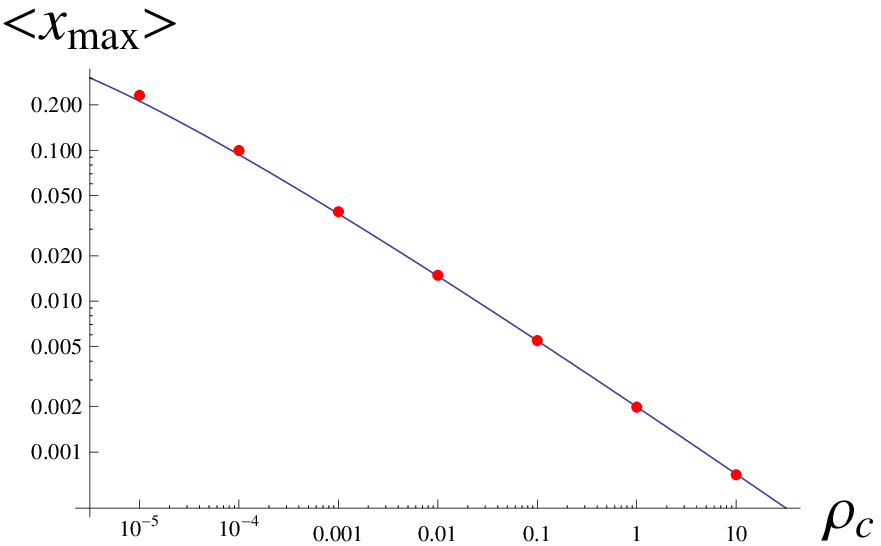}
		\includegraphics[width=\columnwidth]{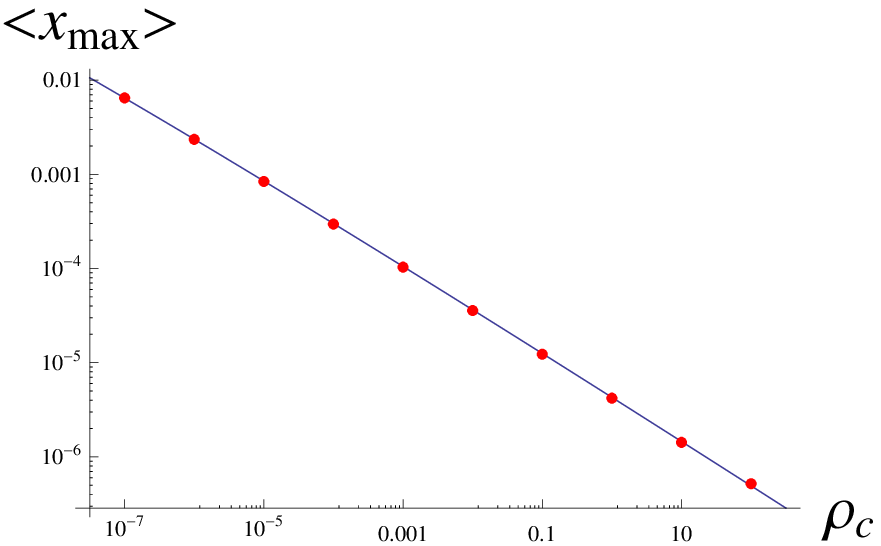}
		\caption{Analytical approximation of $x_{max}$ as a function of $\rho_c$ (blue line) and mean values of numerical simulations (red dots) for $m=10^{-3}$ (upper plot) and $m=1.21\times 10^{-6}$ (lower plot).}
		\label{figxmax}
	\end{center}
\end{figure}

A raw analytical estimate for $N$ can be obtained by assuming that the phase of superdeflation, bounce and superinflation starts at $H=-m$ with $x=0$, and ends at $H=m$. One then finds that
\begin{equation}
\label{max}
x_{max}=\frac{2m}{\sqrt{3\kappa\rho_c}}\ln\left(\frac{2}{m}\sqrt{\frac{\kappa}{3}\rho_c}\right),
\end{equation}
where we have used $\arcsinh\left(\frac{1}{m}\sqrt{\frac{\kappa}{3}\rho_c}\right)\approx\ln\left(\frac{2}{m}\sqrt{\frac{\kappa}{3}\rho_c}\right)$.  
This approximation agrees very well with numerical result as can bee seen in Fig. \ref{figxmax}.
From this, we get the number of e-folds of slow-roll inflation as
\begin{equation}
N=\frac{2}{3}\ln\left(\frac{2}{m}\sqrt{\frac{\kappa}{3}\rho_c}\right)^2.
\label{N2}
\end{equation}

\section{Constraints}
\label{con}

\begin{figure}
	\begin{center}
		\includegraphics[width=\columnwidth]{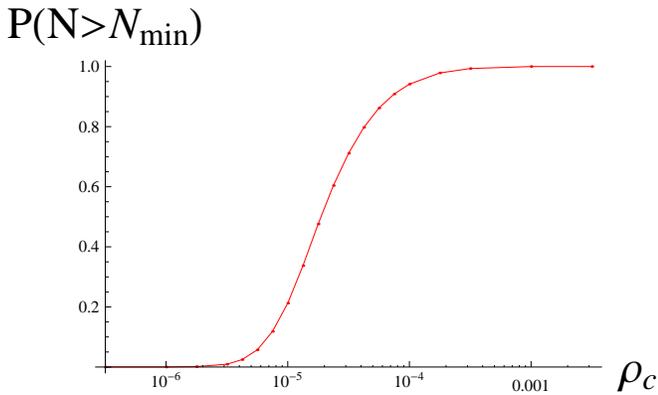}
		\caption{Probability for having more than 65 e-folds of slow-roll inflation, $P(N>65)$, as a function of $\rho_c$.}
		\label{P}
	\end{center}
\end{figure}

%{\renewcommand{\arraystretch}{1.5}
%\renewcommand{\tabcolsep}{0.2cm}
\begin{table}
\begin{tabular}{|c|c|c|}
\hline
$P(N>65)$ & $\rho_c$ & $\gamma$ \\
\hline
$0.5$ & $1.9\times 10^{-5}$ & $6.6$ \\
$0.05$ & $5.4\times 10^{-6}$ & $10.1$ \\
$0.01$ & $3.2\times 10^{-6}$ & $11.9$ \\
\hline
\end{tabular}
\caption{Lower bound on $\rho_c$ and upper bound on $\gamma$, for different minimum required probabilities of a slow-roll inflation longer than 65 e-folds.}
\label{Ptable}
\end{table}

So far, we have used the standard value of $\rho_c$, with a Barbero-Immirzi parameter $\gamma$ assumed to be known from black hole entropy (see, {\it e.g.}, \cite{bh}) in our numerical investigation. By instead taking $\rho_c$ as a free parameter, we can constrain $\rho_c$ and $\gamma$. Previous attempts to constrain $\rho_c$ (see \cite{Mielczarek:2010ga}) from cosmological data were based on $\xmax<1$. However, we have shown that, in all realistic cases, $\xmax$ is much more limited than that.

We can derive an upper limit on $\gamma$ by requiring a large enough probability for a long enough slow-roll inflation. This is again done by assuming the (natural) prior probability
distribution in the prebounce phase that was described previously.
Figure \ref{P} shows $P(N>65)$ as a function of $\rho_c$ and Table \ref{Ptable} gives the constraints on $\rho_c$, and $\gamma$ for different required minimum probabilities for $N>65$. One can also perform an analytical calculation using Eq. (\ref{N2}), leading to $\rho_c>1.6\times 10^{-5}$.

The main results of this analysis are that $\gamma<10.1$ at 95\% confidence level and $\gamma<11.9$ at 99\% confidence level. This is much more stringent than previous cosmological constraints \cite{Mielczarek:2010ga}: $\gamma < 1100$. As the value of $\gamma$ derived form black holes is still controversial, this new bound is clearly meaningful.

\section{Conclusion}

This article establishes a prediction regarding the duration of slow-roll inflation based on holonomy-corrected effective LQC together with a single massive scalar field. The preferred value is $N=145$ e-folds. Values lower than 110 or greeter than 170 are highly improbable. In addition, the value of $x_B$, the square root of the fraction of potential energy at the bounce, is no longer unknown, but is shown to be very close to $3.5\times 10^{-6}$. Finally, the Barbero-Immirzi parameter is now bounded to be smaller than $10-12$ (depending on the confidence level), which is, by far, the best cosmological constraint.

This work should be developed by including other types of matter, by taking into account inverse volume correction, and, in the long run, by trying to use the full LQG theory.

\appendix

\section{Derivation of evolutions in the different phases}
\label{app}
In this appendix, we present the calculations behind the results in section \ref{faces}.\\

\subsection{Oscillations}
These calculations apply to both pre- and post-bounce oscillations.

The first condition of Eqs. (\ref{rhoSmalerThanRhoC1}) and Eqs. (\ref{rhoSmalerThanRhoC4}) ensure that we can approximate Eq. (\ref{Friedman}) by: 
\begin{equation}
H=\pm\sqrt{\frac{\kappa}{3}\rho}.
\end{equation}
In addition, the last condition of Eqs. (\ref{rhoSmalerThanRhoC1}) and Eqs. (\ref{rhoSmalerThanRhoC4})  ensures that we can approximate $x$ and $y$ by oscillating functions with frequency $m$ and varying amplitudes. This, together with Eq. (\ref{rhoxy}), gives Eq. (\ref{prex}).
From this, Eq. (\ref{rhodot}) can be simplified to 
\begin{equation}
\dot{\rho}=\mp 2\sqrt{3\kappa}\ \cos^2(mt+\delta)\ \rho^{3/2},
\end{equation}
which can be integrated to give Eq. (\ref{prerho}) and Eq. (\ref{postrho}).

\subsection{Slow roll}
These calculations apply to both slow-roll deflation and slow-roll inflation.

The last condition of Eqs. (\ref{rhoSmalerThanRhoC2}) and Eqs. (\ref{rhoSmalerThanRhoC3}) ensures that we can approximate Eq. (\ref{rhoxy}) by: 
\begin{equation}
\rho=\rho_c x^2.
\end{equation}
This, together with Eq. (\ref{Friedman}), with the first condition of Eq. (\ref{rhoSmalerThanRhoC2}), and with Eq. (\ref{rhoSmalerThanRhoC3}) gives
\begin{equation}
H=\pm\sqrt{\frac{\kappa}{3}\rho_c}|x|,
\end{equation}
so that the second part of Eq. ({\ref{x}) becomes Eq. (\ref{slowdef}) or Eq. (\ref{slowinf}).

\subsection{Superdeflation, bounce and superinflation}

Without approximations, Eq. (\ref{Friedman}) can be written as
\begin{equation}
H=\pm\sqrt{\frac{\kappa}{3}\rho\left(1-\frac{\rho}{\rho_c}\right)}.
\end{equation}
The second condition of Eq. (\ref{supercond1}) ensures that we can approximate Eq. (\ref{rhoxy}) by 
\begin{equation}
\rho=\rho_c y^2.
\end{equation}
Using the two above equations, Eq. (\ref{rhodot}) can be simplified to
\begin{equation}
\dot{\rho}=\mp 2\sqrt{3\kappa\left(1-\frac{\rho}{\rho_c}\right)}\ \rho^{3/2},
\end{equation}
which can be integrated to give Eq. (\ref{superrho}). It is true both before and after the bounce. Integrating the first part of Eq. (\ref{x}), using the second part of Eq. (\ref{superrho}), gives Eq. (\ref{superx}).

\acknowledgments

This work was supported by the Labex ENIGMASS.

\end{document}